\documentclass[english,letterpaper,aps,preprint]{revtex4}
\usepackage[T1]{fontenc}
\usepackage[latin1]{inputenc}
\usepackage{graphicx}

\makeatletter

\makeatletter



\makeatother

\usepackage{babel}
\makeatother
\begin{document}

\title{Gravitational instability of a dilute fully ionized gas in the presence
of the Dufour effect}

\author{A. Sandoval-Villalbazo$^{1,2}$, A. L. García-Perciante$^{1}$}

\address{$^{1}$Depto. de Matemáticas Aplicadas y Sistemas, Universidad Autónoma
Metropolitana-Cuajimalpa, Av. Pedro Antonio de los Santos No. 84,
México DF, México.\\
$^{2}$Departamento de Física y Matemáticas, Universidad Iberoamericana,
Prolongación Paseo de la Reforma 880, México D. F. 01210, México.}

\date{\today{}}

\begin{abstract}
The gravitational instability of a fully ionized gas is analyzed within
the framework of linear irreversible thermodynamics. In particular,
the presence of a heat flux corresponding to generalized thermodynamic
forces is shown to affect the properties of the dispersion relation
governing the stability of this kind of system in certain problems
of interest. 
\end{abstract}
\maketitle

\section{Introduction}

The study of transport processes in plasmas leads directly to a set
of partial differential equations that describes the evolution of
the local thermodynamic variables relevant to its physical description.
Linear irreversible thermodynamics predicts the coupling of all thermodynamic
 fluxes and forces involving all possible tensors of the same rank
\cite{key-1}. In this context, for multicomponent systems, the problem
of gravitational instability in the presence of a heat flux associated
to a density gradient (Dufour effect) is ought to be examined in order
to address its implications regarding the Jeans instability criterion.
The reciprocal effect, a mass flux due to a temperature gradient (Soret
effect), does not affect the continuity equation for the total density
of the system, and therefore does not alter the Jeans instability
criterion.

While lots of results regarding the effect of dissipation on the Jeans
criterion have been obtained in several works \cite{key-2}, to the
authors' knowledge, this is the first time in which the Dufour effect
is taken into account in calculations regarding gravitational stability.
The numerical values for the transport coefficients used in this work
have been recently obtained by the authors based on the Chapman-Enskog
expansion used in order to solve Boltzmann's equation \cite{key-3}
\cite{key-4}. Similar results have been obtained in terms of several
formalisms, from the works done by Spitzer \cite{key-5}, Braginski
\cite{key-6}, Marshall \cite{key-7} and Balescu \cite{key-8}.

The structure of this work is as follows. In section 2, the basic
hydrodynamic equations are presented including the Dufour flux for
a binary mixture of dilute gases. In section 3 the linearized system
of equations for the fluctuations in the local density $\delta\rho$,
the local temperature $\delta T$ and the hydrodynamic velocity $\delta\vec{u}$
are obtained together with the corresponding dispersion relation associated
to it. Section 4 is devoted to the study of the explicit gravitational
stability criterion for the binary plasma. The last section of this
work is dedicated to a discussion of the results, addressing the conditions
in which the various thermodynamic forces turn out to be relevant
in the study of gravitational stability.

\section{Basic equations and transport coefficients}

Consider a fully ionized system formed by species $i=1,\,2$, $\rho_{1}$
standing for the electron mass density and $\rho_{2}$, corresponding
for the ion mass density. The task to accomplish is to determine the
importance of the various dissipative effects on Jeans instability.
Total mass conservation reads: \begin{equation}
\frac{\partial\rho}{\partial t}+\nabla\cdot\left(\rho\vec{u}\right)=0\label{balancemasa}\end{equation}
 In Eq. (\ref{balancemasa}) $\vec{u}$ corresponds to the hydrodynamic
velocity of the mixture defined by the relation \begin{equation}
\left(\rho_{1}+\rho_{2}\right)\vec{u}=\rho\vec{u}=\rho_{1}\vec{u}_{1}+\rho_{2}\vec{u_{2}}.\label{def1}\end{equation}
The total density of the mixture is given by $\rho=\rho_{1}+\rho_{2}$.
For a quasi-neutral plasma, the number densities are $n_{1}=n_{2}=\frac{n}{2}$,
where $\rho_{i}=m_{i}n_{i}$ and $n=n_{1}+n_{2}$, $m_{1}$and $m_{2}$
being the electron and proton masses, respectively. Also, $\vec{u}_{i}$
represents the average velocity corresponding to species $i$.

The balance equation for linear momentum reads \cite{key-1}: \begin{equation}
\frac{\partial\left(\rho\vec{u}\right)}{\partial t}+\nabla p+\nabla\cdot\left(\rho\vec{u}\vec{u}+\tau^{kl}\right)=-\rho\nabla\varphi\label{momentum}\end{equation}
In Eq. (\ref{momentum}), $p$ represents the local pressure of the
fluid , $\tau^{kl}$ is the viscous contribution to the stress tensor
and $\varphi$ is the gravitational potential which, in Newtonian
mechanics, satisfies the Poisson equation\begin{equation}
\nabla^{2}\varphi=-4\pi G\rho\label{poisson}\end{equation}
Total energy conservation leads to a well known expression for the
evolution of the local temperature of the system $T$, namely \cite{key-1},
\begin{equation}
\rho c_{v}\frac{\partial T}{\partial t}-\left(\frac{\partial\varepsilon}{\partial\rho}\right)_{T}\rho^{2}\nabla\cdot\vec{u}=-\nabla\cdot\left(\vec{J}_{[Q]}\right)-\rho c_{v}\vec{u}\cdot\nabla T-p\nabla\cdot\vec{u}\label{energy1}\end{equation}
where $\rho\varepsilon$ is the internal local energy density. The
heat flux vector $\vec{J}_{[Q]}$ in Eq. (\ref{energy1}) contains
a contribution due to a temperature gradient (Fourier law) and a contribution
associated to a density gradient (Dufour effect): \begin{equation}
\vec{J}_{[Q]}=-\kappa\nabla T-\frac{\mathcal{D}}{2T}\nabla T-\frac{\mathcal{D}}{2\rho}\nabla\rho\label{heatflux}\end{equation}
where $\kappa$ stands for the heat conductivity and $\mathcal{D}$
is the Dufour coefficient. Equations (\ref{balancemasa}), (\ref{momentum})
and (\ref{energy1}), together with their corresponding constitutive
equations relating thermodynamic fluxes and forces, form a complete
set of non-linear partial differential equations which is in general
difficult to work with. Nevertheless, stability analysis around constant
equilibrium values may be readily performed by standard methods. This
will be the subject of the next section.

\section{Linearized transport equations}

A first order perturbative stability analysis can be performed assuming
that any local thermodynamic variable $X$ in this system has a constant
average value $\left\langle X\right\rangle $ and a fluctuation around
it $\delta X$, so that \begin{equation}
X=\left\langle X\right\rangle +\delta X\label{fluc1}\end{equation}
We can now rewrite the system given by Eqs. (\ref{balancemasa}),
(\ref{momentum}) and (\ref{energy1}) in terms of the fluctuations.
For simplicity, $\left\langle \vec{u}\right\rangle =\vec{0}$ is assumed.
The linearized continuity equation becomes: \begin{equation}
\frac{\partial\left(\delta\rho\right)}{\partial t}+\left\langle \rho\right\rangle \delta\theta=0\label{flucmasa}\end{equation}
where $\delta\theta\equiv\nabla\cdot\delta\vec{u}$. Following the
usual approach, the stress tensor in Eq. (\ref{momentum}) is coupled
linearly to the traceless symmetric part of the velocity gradient
through shear viscosity. Thus, the linearized momentum balance equation
becomes: \begin{equation}
\left\langle \rho\right\rangle \frac{\partial\left(\delta\theta\right)}{\partial t}+\frac{3}{5}C_{s}^{2}\nabla^{2}\left(\delta\rho\right)+\left\langle \rho\right\rangle \frac{3C_{s}^{2}}{5T}\nabla^{2}\left(\delta T\right)-D_{v}\nabla^{2}\left(\delta\theta\right)=-4\pi G\left\langle \rho\right\rangle \delta\rho\label{flucmomentum}\end{equation}
where $D_{v}$ =$\frac{4}{3}\eta$, $\eta$ standing for the shear
viscosity measured in international units ($Pa-sec$). In our dilute
(ideal) plasma, bulk viscosity is neglected. We have also used the
fact that, for an ideal gas, $\nabla p=\frac{1}{\gamma}C_{s}^{2}\nabla\rho+\frac{C_{s}^{2}}{\gamma T}\nabla T$
where $C_{s}$ is the adiabatic speed of sound and $\gamma$ is the
heat capacities ratio ($\gamma=5/3$). Therefore, the linearized energy
balance equation is written as: \begin{equation}
\frac{\partial\left(\delta T\right)}{\partial t}-\frac{1}{\left\langle \rho\right\rangle c_{v}}\left(\kappa+\frac{\mathcal{D}}{2\left\langle T\right\rangle }\right)\nabla^{2}\delta T-\frac{1}{\left\langle \rho\right\rangle ^{2}c_{v}}\frac{\mathcal{D}}{2}\nabla^{2}\delta\rho+\frac{2\left\langle T\right\rangle }{3}\delta\theta=0\label{flucT}\end{equation}
Notice that the Dufour coefficient $\mathcal{D}$ vanishes for a simple
fluid \cite{key-7,key-8}, but in principle affects the energy balance
in the binary mixture. For our ideal gas, $c_{v}=\frac{3k}{2m_{2}}$
is the heat capacity measured in $\frac{J}{KgK}$, $k=1.38\times10^{-23}\frac{J}{K}$.
$\kappa$ is measured in $\frac{J}{Kms}$ and $\mathcal{D}$ is given
in $\frac{J}{ms}$. The values of these coefficients for a dilute
plasma have been obtained from plasma kinetic theory \cite{key-4,key-8}.
They can be written as: \begin{equation}
\kappa=\frac{5}{4}\frac{\left\langle n\right\rangle k^{2}\left\langle T\right\rangle }{m_{1}}\left(2.01\tau\right)\label{kappaeff}\end{equation}
 \begin{equation}
\mathcal{D}=\frac{5}{2}\frac{\left\langle n\right\rangle \left(k\left\langle T\right\rangle \right)^{2}}{m_{1}}\left(0.29\tau\right)\label{dufeff}\end{equation}
Here $\tau$ is a characteristic time given by\[
\tau=\frac{4\left(2\pi\right)^{3/2}\sqrt{m_{1}}\epsilon_{0}^{2}\left(kT\right)^{3/2}}{ne^{4}\psi}\]
where $e$ is the electron charge, $\epsilon_{0}$ is he dielectric
constant and $\psi$ is the usual Coulomb logarithm \cite{key-5}.
The shear viscosity coefficient $\eta$ is estimated by using the
Eucken number for an ideal gas, $5/2=\kappa/\eta c_{v}$. We are now
in position to address the importance of dissipation, including the
Dufour effect, on the Jeans instability by means of the analysis of
the corresponding dispersion relation. In order to derive the desired
stability criterion, we perform a Laplace transform in time and a
Fourier transform in space to the system (\ref{flucmasa}-\ref{flucT}).
The corresponding equations become \begin{equation}
s\delta\tilde{\hat{\rho}}+\left\langle \rho\right\rangle \delta\tilde{\hat{\theta}}=\delta\hat{\rho}\left(\vec{q},0\right)\label{s1}\end{equation}
 \begin{equation}
\left(4\pi G\left\langle \rho\right\rangle -\frac{3}{5}C_{s}^{2}q^{2}\right)\delta\tilde{\hat{\rho}}+\left(D_{v}q^{2}+\left\langle \rho\right\rangle s\right)\delta\tilde{\hat{\theta}}-\frac{3\left\langle \rho\right\rangle C_{s}^{2}}{5\left\langle T\right\rangle }q^{2}\delta T=\left\langle \rho\right\rangle \delta\hat{\theta}\left(\vec{q},\,0\right)\label{s2}\end{equation}
 \begin{equation}
\frac{\mathcal{D}}{2\left\langle \rho\right\rangle ^{2}c_{v}}q^{2}\tilde{\hat{\rho}}+\frac{2\left\langle T\right\rangle }{3}\delta\tilde{\hat{\theta}}+\left[s+\frac{1}{\left\langle \rho\right\rangle c_{v}}\left(\kappa+\frac{\mathcal{D}}{2\left\langle T\right\rangle }\right)q^{2}\right]\delta\tilde{\hat{T}}=\delta\hat{T}\left(\vec{q},\,0\right)\label{s3}\end{equation}
Here, the symbol $\tilde{\hat{X}}=\tilde{\hat{X}}\left(\vec{q},\, s\right)$
stands for the successive Laplace-Fourier transforms of the thermodynamic
variable $X$. The wave vector is denoted by $\vec{q}$ and the corresponding
Laplace frequency is $s$.

It is interesting to notice that the first term in Eq. (\ref{s3})
has never been taken into account in the study of the Jeans instability.
The term vanishes in a single component system and may become significant
in the study of transport processes in plasmas, where the difference
of masses between protons and electrons is decisive while constructing
the thermodynamic fluxes and forces \cite{key-3}-\cite{key-4}.

The new dispersion relation reads: \begin{equation}
f\left(s\right)=s^{3}+\alpha s^{2}+\beta s+\gamma=0\label{disp}\end{equation}
 where the coefficients are:\begin{equation}
\alpha=\left(D_{v}+\frac{\kappa}{c_{v}}+\frac{\mathcal{D}}{2c_{v}\left\langle T\right\rangle }\right)\frac{q^{2}}{\left\langle \rho\right\rangle }\label{alpha}\end{equation}
 \begin{equation}
\beta=\left(C_{s}^{2}q^{2}+\frac{D_{v}\mathcal{D}}{2c_{v}\left\langle T\right\rangle \left\langle \rho\right\rangle ^{2}}q^{4}+\frac{D_{v}\kappa}{c_{v}\left\langle \rho\right\rangle ^{2}}q^{4}-4\pi G\left\langle \rho\right\rangle \right)\label{beta}\end{equation}
 \begin{equation}
\gamma=-\frac{4\pi G}{c_{v}}\left(\frac{\mathcal{D}}{2\left\langle T\right\rangle }+\kappa\right)q^{2}+\frac{3C_{s}^{2}\kappa}{5c_{v}\left\langle \rho\right\rangle }q^{4}\label{gamma}\end{equation}

The absence of dissipative effects in Eq. (\ref{disp}) leads directly
to the ordinary Jeans wave number $q_{J}=\sqrt{\frac{4\pi G\left\langle \rho\right\rangle }{C_{s}^{2}}}$.

\section{Stability analysis}

The behavior of the stability condition associated with the polynomial
in Eq. (\ref{disp}) depends significantly on its independent term
which in turn contains dissipative effects. We analyze two different
cases to clarify this point, seeking critical solutions for wave numbers
close to the ordinary Jeans wave number $q_{J}$.

\begin{figure}
\begin{centering}\includegraphics[height=0.5\textwidth]{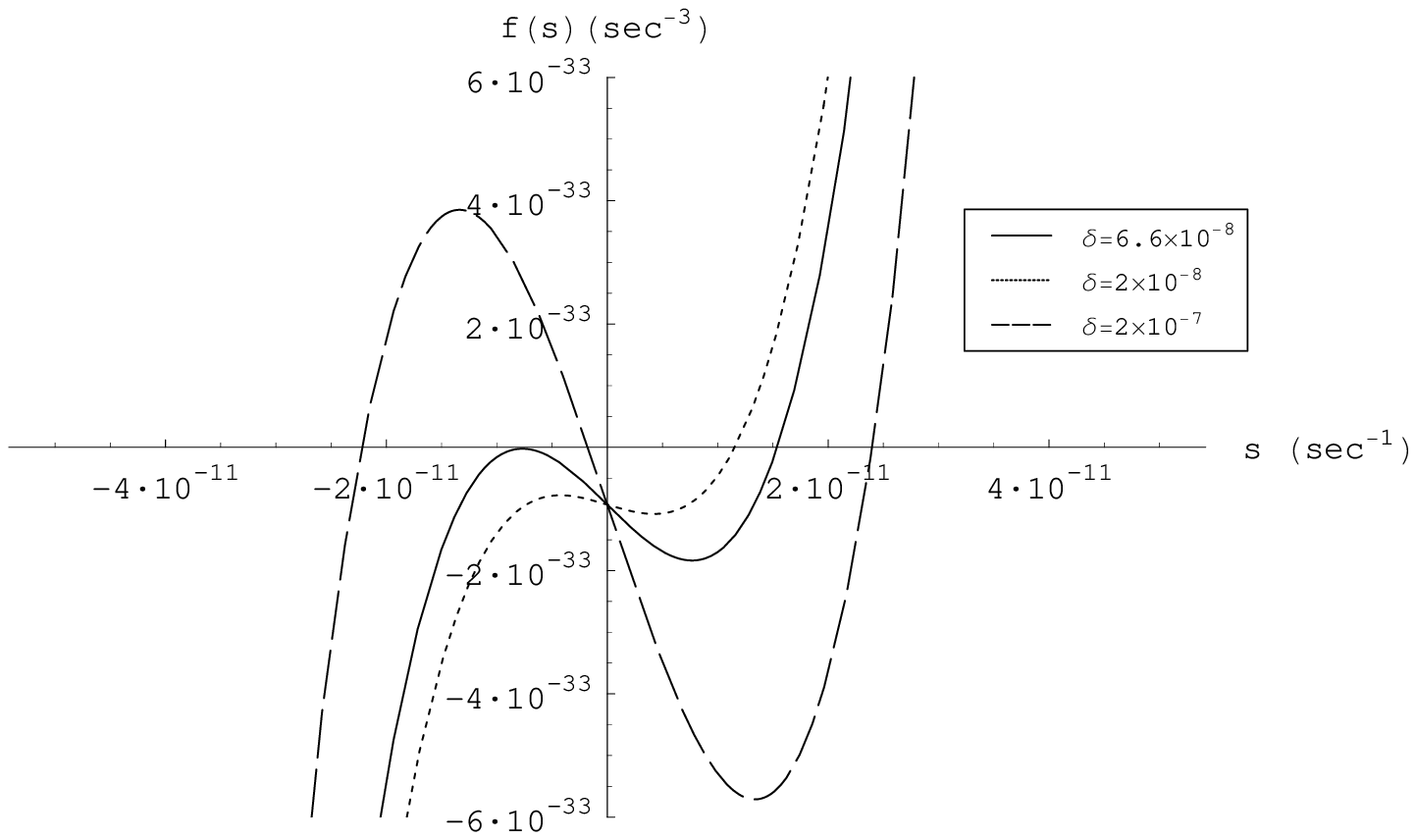}\par\end{centering}

\caption{The function $f(s)$ for values of $q=q_{J}-\delta$ for a number
density $n=10^{21}m^{-3}$ and temperature $T=10^{7}K$.}

\label{fig:1} 
\end{figure}

First consider a system with $n=10^{21}m^{-3}$ and $T=10^{7}K$.
For these values of density and temperature, the cubic function $f\left(s\right)$
in Eq. (\ref{disp}), for a wavenumber $q=q_{J}-\delta$ (where $\delta$
is such that $\delta/q_{J}\ll1$), has in general two critical points
and the independent term is not negligible in the local scale here
considered. In this case, the criterion for stability reduces to finding
the threshold for which the maximum in the negative part of the $s$
axis becomes zero. This can be seen from Fig. \ref{fig:1} which shows
in dotted lines a case for which there is only one real root in the
dispersion relation and one for which three real roots exist. The
threshold that separates the ranges of $q$ for which one obtains
either pure damped or exponentially growing modes is around $\delta=6.6\times10^{-8}$
which is shown in the solid line. %

Analytically, the maximum for $s<0$ is given by\[
s_{-}=-\frac{\alpha}{3}-\frac{\sqrt{\alpha^{2}-3\beta}}{3}\]
 and, since by varying $q$ the whole curve evolves with $f(0)$ virtually
unchanged, the stability threshold is determined by $f(s_{-})=0$.

\begin{figure}
\begin{centering}\includegraphics[height=0.5\textwidth]{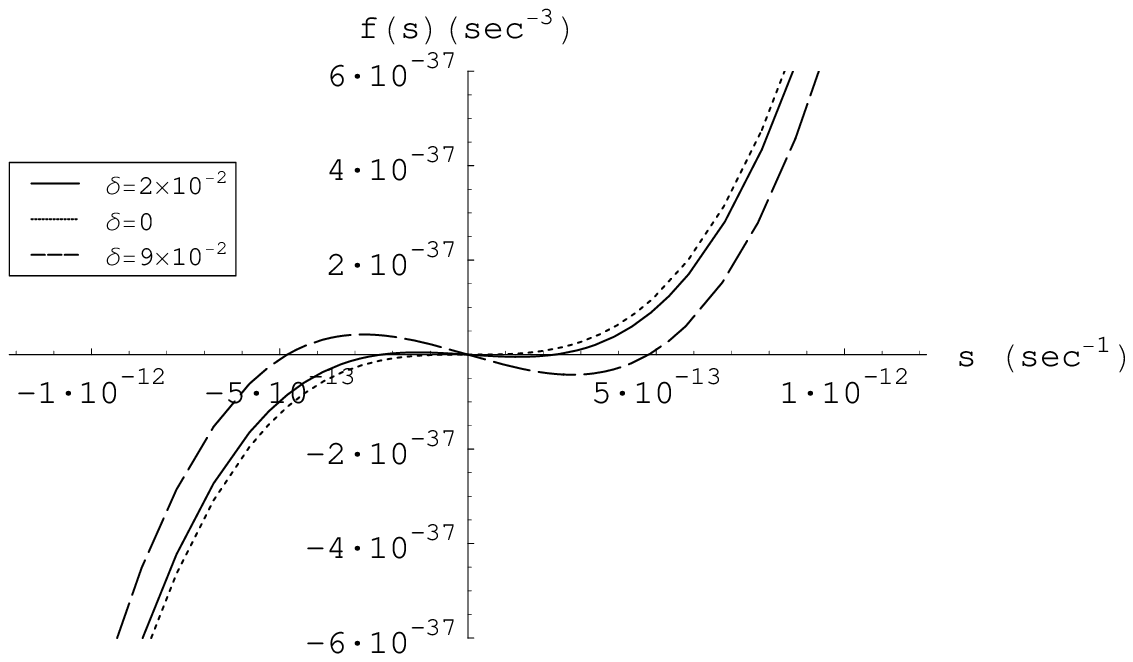}\par\end{centering}

\caption{The function $f(s)$ for values of $q=q_{J}-\delta$ for a number
density $n=10^{12}m^{-3}$ and temperature $T=10^{4}K$.}

\label{fig:2} 
\end{figure}

The second case we shall examine corresponds to lower values of density
and temperature, i. e. $n=10^{12}m^{-3}$ and $T=10^{4}K$. The behavior
in this case is quite different, as shown in Fig. \ref{fig:2}. For
these parameters, both $\alpha$ and the independent term are indeed
negligible and thus, the critical value that gives the wave numbers
is indicated by the disappearance of the two maxima. That is, the
criterion can be found by imposing that the derivative of $f(s)$
has no real roots at all, i. e.\[
\beta>0\]
which, as clearly seen in Eq. (\ref{beta}) simply reduces to the
ordinary Jeans criterion in absence of dissipation.%

\section{Final Remarks}

The problem of the gravitational instability including dissipative
effects has been studied within the framework of linear irreversible
thermodynamics, including cross effects, for a dilute binary plasma.
The dissipative effects enter both the hydrodynamic equations and
the dispersion relation for wave-like solutions of the linearized
system.

The wave number for which perturbations start growing exponentially
in time does not differ much form the standard Jeans wave number and
thus, the response of the system to different fluctuations wavelengths
is similar to the case with no dissipative effects present. However,
the qualitative behavior of the dispersion relation in terms of $q$
is more complicated and has to be explored with care for each system.
Cross effects are present in any multicomponent mixture and are enhanced
by magnetic fields which are observed in many astrophysical systems
\cite{key-4,key-7}. These effects, which are predicted by irreversible
thermodynamics and kinetic theory, are in general not taken into account
while studying gravitational stability and further analysis of them
should be performed in the future.

\begin{acknowledgments}
The authors wish to thank L. S. García-Colín for valuable comments
and fruitful discussions.
\end{acknowledgments}

\end{document}